%% file: main.tex
\documentclass[conference,letter]{IEEEtran}
\usepackage{cite}
\usepackage{amsmath,amssymb,amsfonts}
\usepackage{algorithmic}
\usepackage{graphicx}
\usepackage{textcomp}
\usepackage{xcolor}
\usepackage{url}
\usepackage{hyperref}

\usepackage[caption=false,font=scriptsize]{subfig}

\begin{document}
\title{\huge Tiny Autoscalers for Tiny Workloads: Dynamic CPU Allocation for Serverless Functions}

\author{\IEEEauthorblockN{Yuxuan Zhao}
\IEEEauthorblockA{LIACS, Leiden University \\
y.zhao@liacs.leidenuniv.nl}
\and
\IEEEauthorblockN{Alexandru Uta}
\IEEEauthorblockA{LIACS, Leiden University \\
a.uta@liacs.leidenuniv.nl}
}

\maketitle

\begin{abstract}
In serverless computing, applications are executed under lightweight virtualization and isolation environments, such as containers or micro virtual machines. Typically, their memory allocation is set by the user before deployment. All other resources, such as CPU, are allocated by the provider statically and proportionally to memory allocations. This contributes to either under-utilization or throttling. The former significantly impacts the provider, while the latter impacts the client. To solve this problem and accommodate both clients and providers, a solution is dynamic CPU allocation achieved through autoscaling. Autoscaling has been investigated for long-running applications using history-based techniques and prediction. However, serverless applications are short-running workloads, where such techniques are not well suited.

In this paper we investigate tiny autoscalers and how dynamic CPU allocation techniques perform for short-running serverless workloads. We experiment with Kubernetes as the underlying platform, and implement using its vertical pod autoscaler several dynamic CPU rightsizing techniques. We compare these techniques using state-of-the-art serverless workloads. Our experiments show that dynamic CPU allocation for short-running serverless functions is feasible and can be achieved with lightweight algorithms that offer good performance.

\end{abstract}

\input{intro.tex}

\input{kubernetes.tex}

\input{autoscaling.tex}

\input{experiments.tex}
\input{relwork.tex}
\input{concl.tex}

\section*{Acknowledgments}
The work in this article was in part supported by The Dutch National Science Foundation NWO Veni grant VI.202.195.


\bibliographystyle{IEEEtran}
\bibliography{references_short.bib}

\end{document}

%% file: intro.tex
\section{Introduction}

Serverless computing is gaining significant traction, showing large growths in the past years. Many applications are now running atop serverless offerings, such as machine learning~\cite{carreira2019cirrus,jiang2021towards}, video processing~\cite{jonas2017occupy,fouladi2017encoding}, or analytics~\cite{pu2019shuffling,toader2019graphless,muller2020lambada}. Serverless applications are composed of many short-lived functions running on top of virtualization layers such as containers~\cite{randazzo2019kata} or lightweight virtual machines~\cite{agache2020firecracker}. Throughout the paper, we will refer to both these techniques as \emph{serverless containers}. Typically, serverless functions run for at most 15 minutes and as any other workloads exhibit dynamic resource demands~\cite{gunasekaran2020fifer,bhasi2021kraken}. However, functions are allocated static amounts of CPU time, usually proportional to their memory allocation~\cite{awslambdadoc}. Static CPU allocation in conjunction with dynamic workloads opens up interesting avenues for dynamic autoscaling of serverless containers. Although autoscaling has been investigated for long-running cloud computing applications, it has not been considered yet for dynamic CPU allocation in serverless environments. In this paper we take a step toward rightsizing the CPU allocation of serverless containers dynamically, during runtime, through tiny autoscalers.

In serverless computing, providers run up to thousands~\cite{agache2020firecracker} of serverless containers on a single server to make use of the economy of scale to improve resource utilization. Most of the functions running on these containers are short running and infrequently invoked~\cite{shahrad2020serverless}. However, providers do keep containers around after an invocation has finished to improve cold start latencies for future invocations~\cite{agache2020firecracker,ustiugov2021benchmarking} or predict when functions will be invoked to pre-warm containers~\cite{shahrad2020serverless}. Resource allocators and schedulers take into account also the resources allocated to idle containers, even though they might not be used. Therefore, idle containers that are allocated a large portion of CPU might contribute to the overall under-utilization of a server.     

While running thousands of functions on a server can lead to CPU resource utilization issues, memory utilization is better studied. Users typically select how much memory their functions should be allocated~\cite{awslambdadoc}. While running thousands of these in a single server sounds prohibitive, techniques such as REAP already exist for reducing serverless containers memory footprint~\cite{ustiugov2021benchmarking}. On the other hand, in serverless environments, CPU allocations are static and performed non-transparently to the user, proportionally to the amount of memory allocated. It is thus difficult for a non-technical user to achieve a good CPU allocation for a given application. Although they are short-running, serverless functions exhibit dynamic and non-trivial resource usage, which makes it difficult~\cite{eismann2020sizeless} for their authors to estimate correctly the amount of resources to be requested from the cloud provider.

A solution to this problem is given by rightsizing and autoscaling algorithms. Cloud-based autoscaling has been extensively studied~\cite{ali2012adaptive,iqbal2011adaptive,urgaonkar2008agile,fernandez2014autoscaling,chieu2009dynamic} and many performance-related studies have compared autoscaling algorithms~\cite{ilyushkin2017experimental,bauer2019chamulteon,versluis2018trace}. However, these are based on long-running cloud applications, where historical information is abundantly available for making accurate predictions. In our case, with short running functions that are infrequently invoked, such approaches are not well suited. First, there is little to no historical data available (i.e., in the case of cold starts). It would also be prohibitively expensive to store fine-grained resource usage information for all the functions a provider serves. Second, quick reactions to dynamic resource fluctuations are needed. Third, some of these algorithms are computationally expensive. Running them for thousands of serverless containers at once might be prohibitive. 

Therefore, in this paper we investigate more lightweight rightsizing and autoscaling mechanisms, such as simple-moving average (SMA) and exponential moving average (EMA), inspired from web-based autoscaling techniques~\cite{load-prediction}. We call these tiny autoscalers and compare them with recent methods on container autoscaling~\cite{autoscaling-kth} based on Holt-Winters exponential smoothing~\cite{hw} and long short-term memory (LSTM)~\cite{lstm}. We show the implications of tiny autoscalers and how these can be leveraged by practitioners.

Without loss of generality, we implement tiny autoscalers on top of Kubernetes~\cite{kubernetes}, a container orchestration engine introduced by Google. Similar techniques could be implemented for lightweight virtual machines~\cite{agache2020firecracker} as well using mechanisms such as \emph{Linux cgroups}. Kubernetes makes use of an autoscaling recommender which we override to implement tiny autoscalers. We further compare these with the default Kubernetes autoscaler which was designed for the Google Borg Autopilot~\cite{autopilot}. Our experiments show that the default autoscaler is not well suited for short-running serverless workloads and that significant over- and under-utilization is exhibited through the default Kubernetes autoscaler when running serverless workloads.

Toward showing that dynamic CPU allocation for serverless functions is achievable, our contributions are the following:
\begin{enumerate}
    \item We design, implement, tune and release as open-source tiny autoscalers: lightweight autoscaling mechanisms for serverless workloads. We showcase existing issues in existing state-of-the-art autoscalers that act as barriers in applying them for serverless workloads. (Section~\ref{sec:autoscaling}).
    \item We show empirically that dynamically allocating CPU for short-lived serverless functions through tiny autoscalers is feasible and efficient (Section~\ref{sec:eval}).
    \item We discuss the implications of our results for the design and feasibility of tiny autoscalers for dynamically allocating CPU for serverless workloads (Section~\ref{sec:discussion}).
\end{enumerate}

%% file: kubernetes.tex
\section{System Model}\label{sec:kube}

We make use of the Kubernetes engine to run serverless containers and applications and dynamically alter their allocated CPU during runtime. We describe how Kubernetes works and focus on its default container autoscaler, the Vertical Pod Autoscaler (VPA) Recommender.

\begin{figure}[tp]
    \centering
    \includegraphics[width=0.7\linewidth]{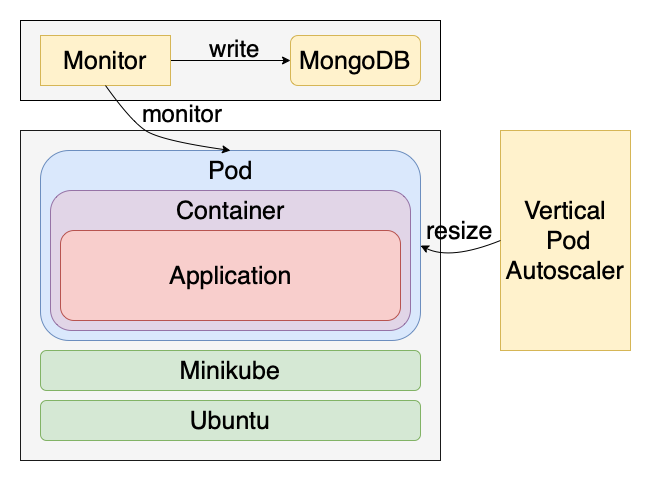}
    \vspace*{-0.4cm}
    \caption{The architecture overview of our system. We deploy a pod containing the application on minikube. The pod is monitored and the monitoring data is stored in a MongoDB database. A VPA is attached to the pod for resizing.}
    \label{fig:architecture}
    \vspace*{-0.4cm}
\end{figure}

The basic architecture of our system is shown in Figure~\ref{fig:architecture}. Our experiment is conducted on a minikube cluster on Ubuntu 20.04. On this minikube cluster, pods are deployed and we configure only one container running per pod. The pods are monitored to record their resource usage in a per-second granularity. The data is collected in a MongoDB database. Furthermore, we enable VPA to provide a mechanism for dynamically resizing the containers resource request. The main aims of VPA are not only reducing the redundant resource waste requested by containers but also reducing the probability of an application in the container being throttled or terminated due to insufficient resources. The VPA primarily consists of three components: recommender, updater, and admission controller. In this article, we mainly focus on the recommender component in VPA. The autoscaling algorithms are integrated into the recommender component and the performance of our algorithms is validated by configuring the VPA with custom autoscaling algorithms. In the following sections, we demonstrate in detail how we monitor the resource usage in Kubernetes and show an overview of VPA architecture~\cite{vpa-flow}.

\subsection{Resource Monitoring}

The resource metrics pipeline~\cite{monitoring-architecture} of Kubernetes reports values through the underlying \emph{Linux cgroups}. These monitoring data are collected by cAdvisor (container advisor)~\cite{cadvisor}, which is a project open-sourced by Google. cAdvisor can collect the information of all the running containers on a machine, including CPU usage, memory usage, etc. Then, cAdvisor is integrated into kubelet~\cite{kubelet}. Kubelet is an agent running on each node in the cluster. Thus, the metrics server~\cite{metrics-server} gets the resource metrics from kubelet and integrates the resource metrics into an API-server, such as the metrics API~\cite{metrics-API}. 

CPU and memory usage of containers are monitored through the \textit{kubernetes.client.CustomObjectsApi} in the Kubernetes python client. Resource requests of pods and containers are read via \textit{kubernetes.client.CoreV1Api}. Resources recommended by VPA are obtained from \textit{kubernetes.client.ApiClient}. Table~\ref{api} lists three data types and their corresponding APIs where we monitor the resource values. All of the resource values mentioned above are monitored every second with fine granularity and stored in a MongoDB database in real-time. The CPU resource is measured in milliCPU (or millicores), a unit of measure introduced by Kubernetes. For example, 100m CPU is equivalent to 0.1 CPU. 

\begin{table}[tp]
    \caption{Data monitoring source.}
    \vspace*{-0.3cm}
    \centering
    \begin{tabular}{c c}
    \hline
    \textbf{Data} & \textbf{API}\\
    \hline
        Resource usage & client.CustomObjectsApi\\
        Resource requests &  client.CoreV1Api\\
        VPA recommendations & client.ApiClient\\
        \hline
    \end{tabular}
    \label{api}
    \vspace*{-0.5cm}
\end{table}

\subsection{Vertical Pod Autoscaler}

The VPA recommender is an essential component in the VPA. It provides the core resource usages (CPU and memory) estimation algorithm which recommends appropriate resource request values for pods in Kubernetes. As a result, the containers would resize according to the recommended resource values. Therefore, the quality of the recommendation algorithm largely determines the quality of the container resizing. An application would be throttled if it exceeds the specified CPU limit of the container and gets terminated if its memory exceeds the limited amount of the container.

The VPA is mainly composed of recommender, updater, and admission controller. The recommender watches all the pods in the Kubernetes cluster and calculates the recommendation values for each pod. The recommendation algorithm is included in section~\ref{vpa-recom-algo}. The recommender reads the pod metrics from Prometheus~\cite{prometheus} regularly. The updater is responsible for updating pods according to pod recommendations. Currently, the VPA updater only supports evicting old pods before recreating a new pod with the recommended resources. This means the service will be disrupted when a pod needs an update. In-place update~\cite{in-place} was proposed, but it is still under development. In this paper we turn off the updating mechanism and only consider the recommendations, thus not disrupting containers. This is sufficient to showcase the contributions of this article. 

%% file: autoscaling.tex
\section{Autoscaling Methods}\label{sec:autoscaling}

We discuss autoscaling methods and explain how one can adapt them to the problem of dynamically rightsizing the CPU allocation of serverless functions. We identify drawbacks for these techniques and show how we overcome these.

We first introduce the resource estimation algorithm in the original Kubernetes VPA. The recommendation algorithm currently used in vertical pod autoscaler is deeply inspired by the moving window recommender in Google Borg Autopilot~\cite{autopilot}. Then we discuss the autoscaling strategy proposed in~\cite{autoscaling-kth}, which primarily applies Holt-Winters exponential smoothing (HW)~\cite{hw} and Long Short-Term Memory (LSTM)~\cite{lstm} algorithms to predict the future resource demands. Lastly, we present a resource estimation algorithm largely based on the ideas in CPU usage prediction models in web-based system~\cite{load-prediction} and bring forth new ideas of how to adapt these algorithms to the problem at hand. We improve the algorithms to adapt to the scenario of serverless function CPU resizing. 


Table~\ref{embed} gives a summary of the methods used in this paper. Besides our newly implemented algorithms, the original VPA recommendation algorithm is already implemented into vertical pod autoscaler. As for HW and LSTM autoscaling strategy proposed in~\cite{autoscaling-kth}, they are not integrated into the vertical pod autoscaler component by their authors and it is outside the scope of this paper to implement them in the Kubernetes VPA. We instead use emulation to compare these algorithms with tiny autoscalers.

\begin{table}[tp]
    \caption{Methods considered and compared in this article.}
    \vspace*{-0.3cm}
    \centering
    \begin{tabular}{c c}
    \hline
    \textbf{Methods} & \textbf{Integrated with Kubernetes VPA}\\
    \hline
        VPA recommender & Yes (existing)\\
        HW recommender &  No, emulation\\
        LSTM recommender & No, emulation\\
        SMA recommender & Yes (our contribution)\\
        EMA recommender & Yes (our contribution)\\
        \hline
    \end{tabular}
    \label{embed}
    \vspace*{-0.5cm}
\end{table}

\subsection{Kubernetes VPA Recommender}\label{vpa-recom-algo}

The recommender of the VPA mainly borrows the ideas of the moving window recommender in Google Borg Autopilot~\cite{autopilot}. The VPA recommender creates a decaying histogram object to store the CPU usage for every container. The recommender acquires the resource usage of all pods from Prometheus~\cite{prometheus} regularly and writes the resource usage of containers into a maintained corresponding decaying histogram. The decaying histogram is composed of multiple buckets, which are used to store the weight of resource usage. The size of buckets in a decaying histogram grows exponentially with a ratio of 1.05. The first bucket stores the weights of resource usage in the range of $[0, firstBucketSize)$. Since the bucket size grows exponentially, the size of the $nth$ bucket follows $firstBucketSize * ratio^{n-1}$.
The weight of every resource usage is stored in the bucket where the resource usage falls between the bucket boundaries. As time goes by, usage weight decreases as well. If the default \emph{$HalfLife$} is set to 24h, the weight of the past 24 hours before will be halved.

In terms of recommendation, the VPA recommender uses three values: target value, lower bound value, and upper bound value. They are the starting values (left boundaries) of the bucket where the total weight of that bucket and the former buckets arrives at $0.9*totalWeight$, $0.5*totalWeight$, and $0.95*totalWeight$ for the first time, correspondingly. Furthermore, VPA recommender applies a confidence multiplier to lower bound value and upper bound value. 
The VPA evicts the pod as soon as its request value is beyond the range of upper bound and lower bound, then creates a pod with the current recommended value as its request. 

\noindent\fbox{%
    \parbox{\linewidth}{%
        \textbf{Kubernetes VPA recommender drawback:} The Kubernetes default VPA recommender, based on Google Borg Autopilot~\cite{autopilot}, is not sensitive to short, sudden changes in workloads, and is more suited for longer running workloads. Serverless workloads are short running with sudden bursts in CPU usage.
    }%
}

\subsection{HW and LSTM Recommender}
Wang et al. proposed an autoscaling mechanism~\cite{autoscaling-kth} which applies Holt-Winters exponential smoothing (HW)~\cite{hw} and Long Short-Term Memory (LSTM)~\cite{lstm} algorithms to increase the CPU utilization of Kubernetes containers. Their autoscaler takes target value, lower bound value, and upper bound value from HW and LSTM models as input and supplies 120 millicores as the error buffer. 
Their algorithm will give a new recommended value when the current CPU request is out of the range of bounds. They preset two values to avoid unnecessary rescaling. One is the rescaling cool-down value, the other is the minimum change check value. This means it will rescale only after a cool-down time has passed since the last rescaling. In addition, the difference between the value of the current request and a new request must be more than minimum change check value. HW and LSTM recommenders are implemented in Python with \textit{Statsmodels} and \textit{Keras}. For HW recommender, the prediction model refits when each new observation is collected. For LSTM recommender, the prediction model retrains every season (several observations compose a season). At least two season data are needed to generate the future CPU usage prediction from HW and LSTM recommenders because these two models need to be initialized at the beginning. That explains why there is a fixed recommendation value in cold starts for both HW and LSTM recommenders (See Section~\ref{Dynamic CPU Allocation for Cold Starts}).

\textbf{Emulation.} Unfortunately, this autoscaling mechanism is not integrated into the VPA. Thus in this article the performance of HW and LSTM is not benchmarked in real-time but rather in emulation. We initially run the experiments without any VPA and collect the CPU usage data. Then we train the HW and LSTM models using a subset of the initial data. The season length is set to one minute and as training data we use two seasons. Using any shorter seasons will result in very poor performance. The input data are fed into these two models to refit or retrain the models and get the predicted values of future CPU usage. The final output of these models is the recommended value for the future CPU demands. 

The major downside of the HW and LSTM recommenders is that they need significant amount of training data to perform well. In short running serverless workloads, this is either difficult to get since the functions are very short, or very expensive to store since providers are running extremely large numbers of different functions. Keeping utilization data for all of them will be prohibitive. Hence, The HW and LSTM recommenders are expensive for serverless containers. \\

\noindent\fbox{%
    \parbox{\linewidth}{%
        \textbf{LSTM and HW drawback:} ML-based recommenders need significant amount of data for training, which might not be available in serverless scenarios. Moreover, ML-based recommenders are computationally expensive. 
    }%
}

\subsection{Tiny Autoscalers: SMA-and EMA-based}
Andreolini et al.~\cite{load-prediction} proposed load prediction models for CPU usage in web-based systems. Since these models are lightweight to apply in terms of computational complexity and do not need much training data, in this paper we investigate if they are suitable for short-lived serverless workloads as tiny autoscalers. We innovate these algorithms while inheriting the ideas behind their prediction models. We adapt the models and implement them in the Kubernetes VPA. 

\begin{figure}[tp]
    \centering
    \includegraphics[width=0.35\linewidth]{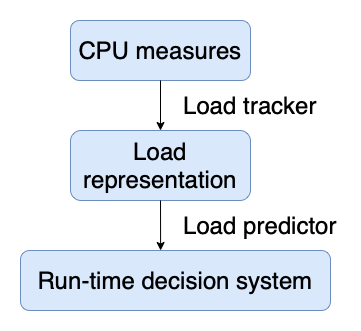}
     \vspace*{-0.4cm}
        \caption{The basic framework of load prediction models~\cite{load-prediction}. The actual CPU usage are processed by two types of load trackers and get CPU usage representations, then make prediction for future CPU demands according to the representations.}
    \label{framework}
    \vspace*{-0.6cm}
\end{figure}

Andreolini et al. demonstrate in their paper~\cite{load-prediction} that it is infeasible to predict future load well using the raw CPU usage measures. They design load trackers, two linear functions to smooth the trend of CPU usage, representing the CPU load behavior of the system (see Section~\ref{load-tracker}). Then, they make predictions of CPU usage through load representations processed by load trackers. The basic framework of the load prediction models is depicted in Figure~\ref{framework}. 

\subsubsection{Load trackers}\label{load-tracker}
As Figure~\ref{framework} shows, the load prediction model has a two-step approach. In the first phase of the models, we still apply two linear load tracker functions presented in the original paper~\cite{load-prediction}, which are simple moving average (SMA) load tracker and exponential moving average (EMA) load tracker. Given a CPU usage value $s_i$ measured at time $t_i$ and previously sampled $n$ CPU usage values, they compose a set $S_{n}\left(t_{i}\right)=\left(s_{i-n}, \ldots, s_{i}\right)$. The load tracker function is defined as $LT\left(S_{n}\left(t_{i}\right)\right): \mathbb{R}^{n+1} \rightarrow \mathbb{R}$, where the $S_{n}\left(t_{i}\right)$ is the input of the load tracker function and it returns a representation $l_i$ to represent the set of $S_{n}\left(t_{i}\right)$ at time $t_i$. Simple moving average (SMA) is the unweighted average of $n+1$ CPU usage values in the set $S_{n}\left(t_{i}\right)$, the weights assigned to each observation are the same. So the SMA-based load tracker function at time $t_i$ is defined as:
$
SMA\left(S_{n}\left(t_{i}\right)\right)=\frac{\displaystyle\sum_{i-n \leq j\leq i} s_j}{n+1}.
$
The problem of the SMA load tracker in theory is that it will involve a delay when it represents the workload trend, especially if the size of $S_{n}\left(t_{i}\right)$ is large. As opposed to SMA, exponential moving average (EMA) load tracker function can decrease the delay effect well in theory. Exponential moving average (EMA) is the weighted average of $n+1$ CPU usage values in the set $S_{n}\left(t_{i}\right)$ and the weights of observations are exponentially decreasing. Thus, the EMA-based load tracker function at time $t_i$ is defined as:
\begin{equation*}
\begin{split}
&EMA\left(S_{n}\left(t_{i}\right)\right) = \\
  &=\begin{cases}
    \frac{\displaystyle\sum_{0 \leq j\leq n} s_j}{n+1} & \quad \text{if } i \leq n,\\
    \alpha * s_{i}+(1-\alpha)*\operatorname{EMA}\left(S_{n}\left(t_{i-1}\right)\right)  & \quad \text{if } i > n.
  \end{cases}
\end{split}
\end{equation*}
The parameter $\alpha$ is called the smoothing constant. We conform with the constant value in the paper~\cite{load-prediction} and set the smoothing constant $\alpha = \frac{2}{n+1}$. For the load tracker based on EMA at time $t_i$, the recent observations contribute more to the representation $l_i$ than the older observations due to the decaying weights.

\subsubsection{Load predictor}
In the second phase of the models, we adapt and innovate the load prediction in the paper~\cite{load-prediction} according to our usage scenarios. The load predictor function is defined as $LP\left(L_{q}\left(t_{i}\right)\right): \mathbb{R}^{n+1} \rightarrow \mathbb{R}$, where $L_{q}\left(t_{i}\right)=\left(l_{i-q}, \ldots, l_{i}\right)$ is a set of $q+1$ representations obtained from load tracker function. $LP\left(L_{q}\left(t_{i}\right)\right)$ returns the predicted future CPU usage value. $LP\left(L_{q}\left(t_{i}\right)\right) = Max(\beta*S/EMA\left(S_{n}\left(t_{i}\right)\right),\quad m *(i+k)+a)$, where   $m=\frac{l_{i}-l_{i-q}}{q}$, and $a=l_{i-q}-m * (i-q)$.

\noindent\fbox{%
    \parbox{\linewidth}{%
        \textbf{EMA and SMA drawback:} While EMA and SMA algorithms are lightweight and effective in following the resource demand, we experimentally found that EMA and SMA need to be tuned for the workloads they target. As a consequence, we adapt EMA and SMA to work efficiently on serverless workloads. In Section~\ref{sec:eval:tuning} we show how our tuned versions perform.
    }%
} \\

The fundamental idea behind our tuning and modification for the load predictor is an heuristic which will always try to slightly over-provision than under-provision, so that the client serverless function will not be significantly throttled. Thus, in our prediction algorithm, we introduced $\beta$, a constant to multiply with the former average of recent CPU usage values. We then take the maximum between this term and the original predicted value. This mechanism is a \emph{bottoming} mechanism so that the prediction does not drop very rapidly.

\begin{table}[tp]
    \caption{Serverless workloads used in our experiments.}
    \vspace*{-0.2cm}
    \centering
    \begin{tabular}{ccc}
    \hline
        \textbf{Name} & \textbf{Input Size} & \textbf{Runtime} \\ \hline
        image\_rotate & 10,000 Images &  ~4 minutes\\
        image\_rotate\_shorter & 1,000 Images & ~15 seconds\\
        lr\_training\_default & 5,700MB CSV & ~10 minutes\\
        video\_processing\_17m & 17MB Video & ~15 seconds\\
        video\_processing\_67m & 67MB Video & ~1.5 minutes\\
        video\_processing\_127m & 127MB Video & ~4 minutes\\
        
         \hline
    \end{tabular}
    \label{workloads}
    \vspace*{-0.3cm}
\end{table}

\begin{table}[t]
    \centering
    \caption{Metrics interval influence on Default VPA over-provisioning.}
   \vspace*{-0.2cm}
    \begin{tabular}{cc}
    \hline
        \textbf{Metrics Interval} & \textbf{Average Slack (millicores)} \\ \hline
        1 minute & 426.23\\
        1 second & 328.23 \\
         \hline
    \end{tabular}
    \label{tuning default}
    \vspace*{-0.6cm}
\end{table}

Since in the original formula we have the term $m=\frac{l_{i}-l_{i-q}}{q}$ in the prediction, $m$ will be close to 0 when the CPU load tends to be flat, which will lead to a \emph{cliff} on the predicted values. Thus, we introduce a novel \emph{bottoming} mechanism to the EMA and SMA algorithms. The adoption of the bottoming mechanism prevents the predicted resource usage value from rapid and unexpected decline. Additionally, the linear extrapolation prediction, according to recent representations from load tracker, ensures the predicted value to rise abruptly when CPU usage is at peak. SMA-based and EMA-based recommenders have the advantage of their low computational complexity, so they are suitable for tiny autoscalers for serverless functions while keeping a promising prediction result. 

In terms of update policy, we follow the design in the VPA keeping the lower bound and upper bound for a recommendation value. However, what is different from VPA is that we revise the calculation methods of lower bound and upper bound. In VPA, the lower bound and upper bound are calculated as the starting values of the bucket where its accumulated weight achieves 50\% and 95\% of total weight. In SMA- and EMA-based recommender, the upper bound and lower bound follow the trend of predicted value and with a multiplier as in the VPA. Thus, the lower bound and upper bound will converge to the predicted value as time goes by in the same as the default VPA. When the request value of the pod is out of the range of lower bound and upper bound, the pod will be updated with a new request value that is the same as the recommended value at that time.

Two parameters are needed in both SMA-based and EMA-based recommenders, which are the size of the load tracker and the number of the load trackers. The size of the load tracker represents how many observations are in a load tracker to calculate the (un)weighted average. The number of load trackers indicates how many representations $li$ obtained from the load tracker function are used to make an extrapolation prediction. For example, we use EMA5-3 to represent EMA method with load tracker of size 5 and with 3 load trackers. 

%% file: experiments.tex
\section{Experimental Evaluation}\label{sec:eval}

We investigate empirically what are implications of dynamically autoscaling the CPU allocations of serverless containers. We explain our experimental setup, tune the default Kubernetes VPA, show our results tuning EMA- and SMA-based autoscalers. Finally, we show what are the implications of dynamically rightsizing CPU allocations using 4 different autoscalers for serverless workloads in both cold and warm starts. We discuss the practical implications of our results.

\begin{figure}[tp]
    \centering
    \includegraphics[width=0.95\linewidth]{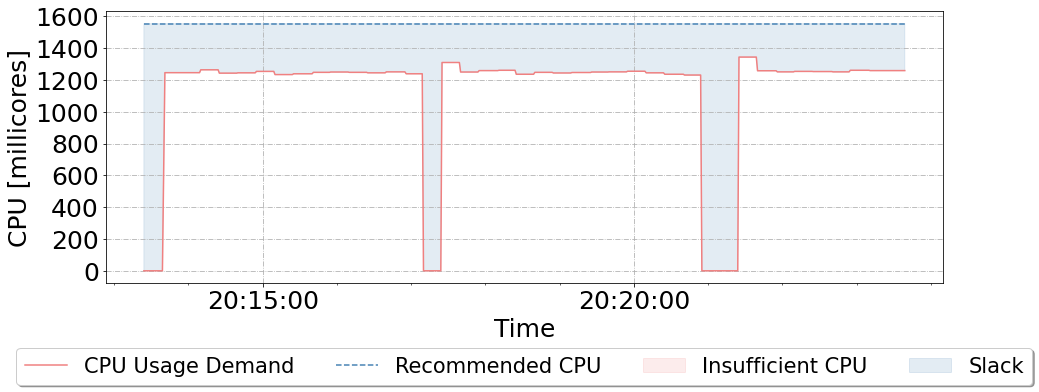}
    \vspace*{-0.3cm}
        \caption{Result of video\_processing\_127m with default VPA (metrics interval 1 minute). On this workload, the default VPA recommendation stays unchanged and cannot react with changes of the actual CPU usage.}
    \label{127m_default}
    \vspace*{-0.4cm}
\end{figure}

\begin{figure}[tp]
    \centering
    \includegraphics[width=0.95\linewidth]{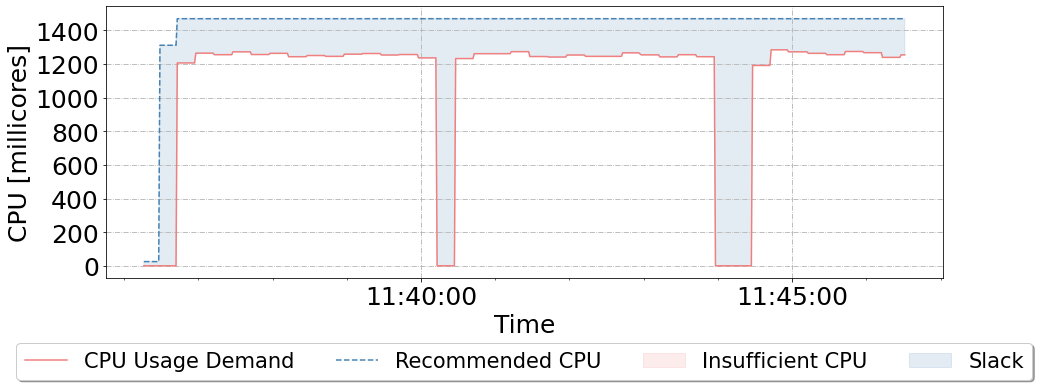}
    \vspace*{-0.3cm}
    \caption{Result of video\_processing\_127m with default VPA (metrics interval 1 second). The default VPA with metrics interval as 1 second can follow the trend of the actual CPU usage at the beginning but stays unchanged in the following warm starts.}
    \label{127m_default1s}
    \vspace*{-0.5cm}
\end{figure}

\subsection{Experiment Setup}
In this paper, we investigate the performance of several CPU autoscaling mechanisms on serverless workloads. These CPU autoscaling mechanisms include the default VPA in Kubernetes, simple moving average (SMA) and exponential moving average (EMA) which are inspired from web-based load prediction models~\cite{load-prediction}, and container autoscaling methods~\cite{autoscaling-kth} based on Holt-Winters exponential smoothing (HW) and long short-term memory (LSTM). The performance of these mechanisms are measured by two metrics: average slack and average insufficient CPU. Slack is the amount of CPU usage that is over-provisioned, while insufficient CPU usage refers to the amount of CPU usage that is under-provisioned. The metrics are defined in detail by Wang et al.~\cite{autoscaling-kth} and can be viewed as a simplification of the extensive set of metrics defined by Ilyushkin et al.~\cite{ilyushkin2017experimental}.


To evaluate these autoscalers we use three workloads from the vHive~\cite{ustiugov2021benchmarking} archive. By using different input sizes which shorten or lengthen their running time we reach a total of six serverless workloads. According to the Microsoft investigation of their own serverless workloads~\cite{shahrad2020serverless}, our proposed workloads are realistic in terms of runtime. Table~\ref{workloads} shows the serverless workloads evaluated in this paper and their runtime.

We run all our experiments under Ubuntu 20.04 and minikube v1.24.0. Our servers are an on-prem machine with 8 cores and 16\,GB RAM and \textit{i3.metal} AWS EC2 virtual machine, which has 72 cores and 512\,GB RAM. 

\subsection{Tuning the Kubernetes Default VPA Autoscaler}

The default Kubernetes VPA Autoscaler makes recommendations at fixed-time intervals. The default interval is 1 minute. This is insufficient for serverless workloads which could be much shorter than 1 minute as shown by Shahrad et al.~\cite{shahrad2020serverless}. To improve the default Kubernetes VPA autoscaler we have tweaked this value to 1 second. We show the difference between the two time intervals in the following experiment. 

To evaluate the performance of the default VPA autoscaler, we run all the workloads repeatedly. The first run of a workload emulates a cold start, while the subsequent runs emulate warm starts of the serverless functions. Figures~\ref{127m_default} and~\ref{127m_default1s} plot our results for the video processing workload. Interpreting these results, one could notice that the VPA with 1 second interval acts faster at the beginning because the first recommendation from autoscaler with 1 minute metrics interval only can be caught after 1 minute. Although both of them cannot follow the trend of CPU usage fluctuations, the autoscaler with 1 second metrics interval has less slack than autoscaler with 1 minute metrics interval. Table~\ref{tuning default} shows the average slack of default VPA autoscaler with different metrics intervals.
\\

\noindent\fbox{%
    \parbox{\linewidth}{%
        \textbf{Conclusion-1:} The default Kubernetes VPA cannot follow the actual CPU usage fluctuations closely. With 1 second metrics interval it exhibits less slack than for 1 minute metrics interval. Short-running serverless workloads need different autoscalers (Figures~\ref{127m_default}, \ref{127m_default1s}, Table~\ref{tuning default}).
    }%
}

\begin{table}[t]
    \caption{Comparing average slack and average insufficient CPU of ema5-3 before and after our tuning and adaptation.}
    \vspace*{-0.3cm}
    \centering
    \begin{tabular}{ccc}
    \hline
        \textbf{Tuning} & \textbf{Avg. Slack (millicores)} & \textbf{Avg. Insuf. (millicores)} \\ \hline
        Before tuning &5.16 &288.75\\
        After tuning &134.81 &66.89\\
         \hline
    \end{tabular}
    \label{tab:before_and_after_adaption}
    \vspace*{-0.6cm}
\end{table}

\subsection{Building Tiny Autoscalers: Tuning EMA and SMA}\label{sec:eval:tuning}

In Section~\ref{sec:autoscaling} we introduced the SMA and EMA methods for web-based workload CPU usage prediction~\cite{load-prediction}. We have found that in their original form they do not perform well for serverless workloads, exhibiting significant under-provisioning. Thus, in this paper, we adapt these methods to serverless workloads as explained in Section~\ref{sec:autoscaling}. To show the under-provisioning behavior more clearly, in this section we use a different workload, namely the YSCB~\cite{ycsb} key-value store benchmark running on top of the Redis key-value store. The VPA we implemented monitors the Redis Kubernetes Pods and adapts their CPU dynamically.

\begin{figure}[tp]
    \centering
    \includegraphics[width=0.99\linewidth]{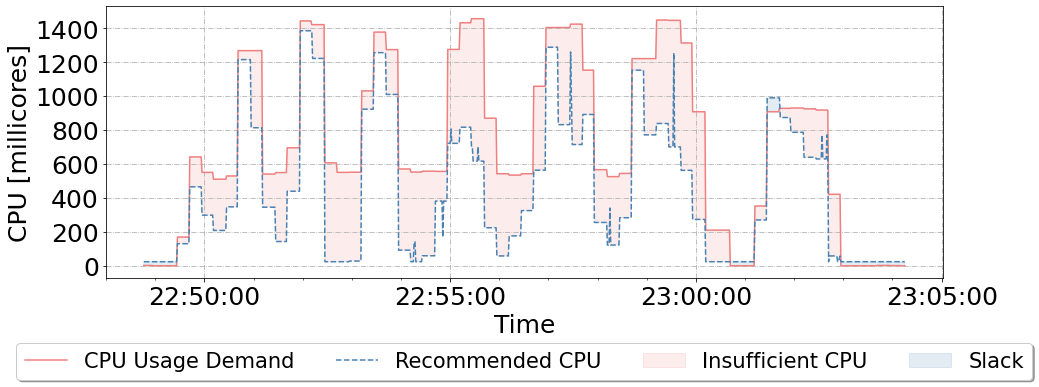}
    \vspace*{-0.7cm}
        \caption{Default ema5-3 dynamic CPU allocation (before tuning). Notice how the default EMA method consistently allocates insufficient CPU.}
    \label{fig:before_tuning}
    \vspace*{-0.4cm}
\end{figure}

\begin{figure}[tp]
    \centering
    \includegraphics[width=0.99\linewidth]{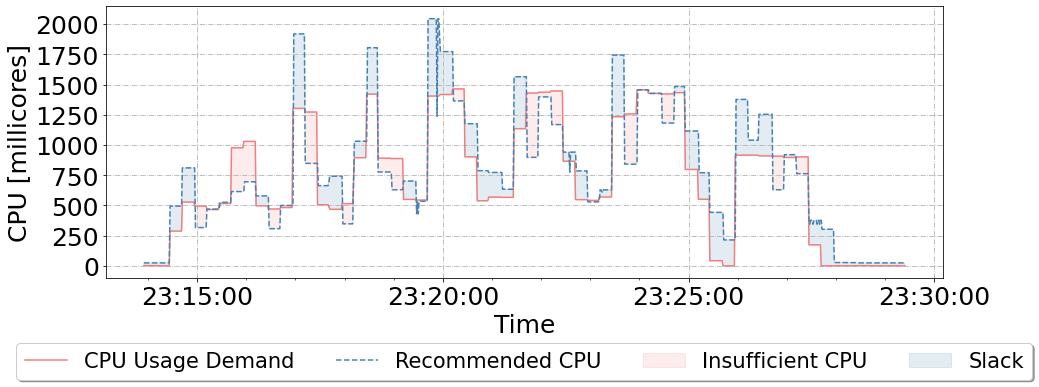}
    \vspace*{-0.7cm}
        \caption{Dynamic CPU allocation using ema5-3 including our tuning defined in Section~\ref{sec:autoscaling}. Notice how the tuned EMA version has a much more balanced CPU allocation.}
    \label{fig:after_tuning}
    \vspace*{-0.6cm}
\end{figure}

As Figure~\ref{fig:before_tuning} shows, the prediction of the unaltered original method can react with the change of CPU usage trend very fast, not only for increasing CPU but also for decreasing CPU. However, as the figure shows, this encompasses significant amounts of insufficient CPU (i.e., under-provisioning). This results in the application being slowed down. To overcome this disadvantage, we introduce a \emph{bottoming} mechanism to the algorithms. We involve unweighted or weighted average of recent CPU usage values to prevent the prediction from rapid decline unexpectedly and unreasonably. As Figure~\ref{fig:after_tuning} indicates, after our adaptation and tuning, the prediction is improved when the actual CPU usage decreases suddenly. Table~\ref{tab:before_and_after_adaption} compares average slack and average insufficient CPU of ema5-3 before and after our adaptation.

In this section we present only the results of the ema5-3 method due to space limitations, but the results we have are consistent over all EMA and SMA methods, e.g., ema10-5, ema3-2, sma3-2, sma5-3. In the latter experiments presented in this section we show overall results using all these methods which perform better than the default Kubernetes VPA and the ML-based methods of LSTM and HW.
\\

\noindent\fbox{%
    \parbox{\linewidth}{%
        \textbf{Conclusion-2:} The tiny autoscalers can predict the actual CPU usage closely after our tuning. Our adaptation is able to leverage much less under-provisioning, offering the application better overall CPU performance (Figures~\ref{fig:before_tuning}, \ref{fig:after_tuning}, Table~\ref{tab:before_and_after_adaption}).
    }%
}

\subsection{Dynamic CPU Allocation for Cold Starts}\label{Dynamic CPU Allocation for Cold Starts}

One very important area of research related to serverless systems are cold starts. They refer to the first start of a serverless function on a given server, when the underlying subsystem is not pre-warmed (e.g., microVMs are not yet booted, runtimes are not loaded etc.). The first start might refer to either the first invocation of a function after its creation, or the first start of a function on a specific server. The behavior would be similar in both cases.

For scheduling containers in serverless workloads, cold starts are important because schedulers that make use of historic information, such as the default Kubernetes VPA, LSTM or HW discussed earlier in this paper, do not have any historical information. Even though a function might have been run previously on a different server, we assume that for cloud providers keeping fine-grained scheduling and resource usage information is prohibitive, as providers are likely to run millions of such functions every hour~\cite{shahrad2020serverless}. Under such auspices, it is important to know how autoscalers fair for the first time when scheduling a specific function.

\begin{figure}[tp]
    \centering
    \includegraphics[width=0.95\linewidth]{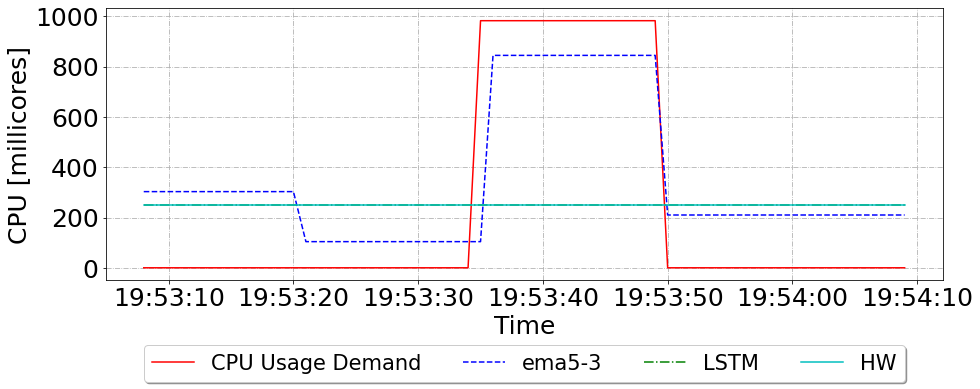}
    \vspace*{-0.4cm}
    \caption{Cold starts for HW and LSTM compared to ema5-3 when running image\_rotate\_shorter workload. Due to lack of historical information and training, LSTM and HW cannot allocate sufficient CPU. LSTM and HW curves overlap.}
    \label{fig:cold_image}
    \vspace*{-0.4cm}
\end{figure}

\begin{figure}[tp]
    \centering
    \includegraphics[width=0.95\linewidth]{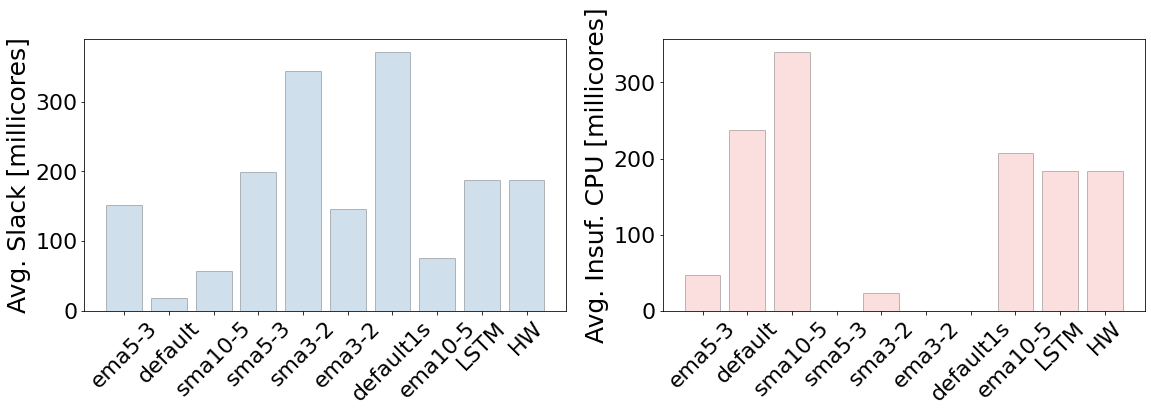}
    \vspace*{-0.4cm}
    \caption{Comparing the average slack and insufficient CPU among all the methods in this paper when running image\_rotate\_shorter workload in the cold start. Due to the training process of LSTM and HW, they do not perform well in both slack and insufficient CPU as a result of presenting a preset value.}
    \label{fig:cold_image_bar}
    \vspace*{-0.6cm}
\end{figure}

\begin{figure}[tp]
    \centering
    \includegraphics[width=0.95\linewidth]{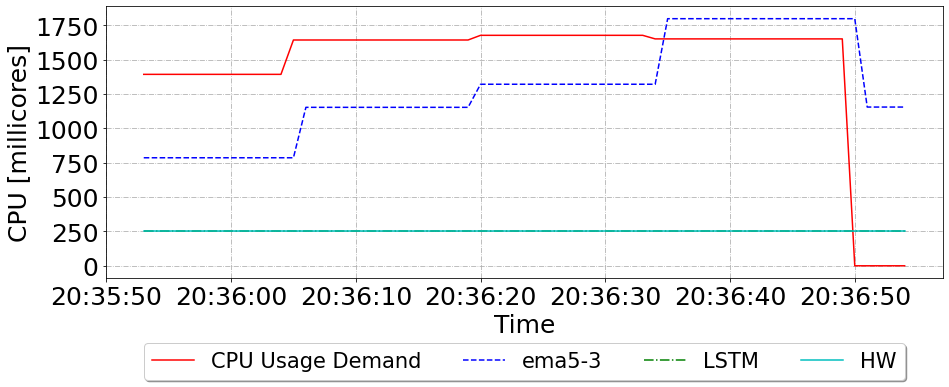}
    \vspace*{-0.4cm}
    \caption{Cold starts for HW and LSTM compared to ema5-3 when running video\_processing\_17m workload in the first run. Due to lack of historical information and training, LSTM and HW cannot allocate sufficient CPU. LSTM and HW curves overlap}
    \label{fig:cold_video}
    \vspace*{-0.4cm}
\end{figure}

\begin{figure}[tp]
    \centering
    \includegraphics[width=0.95\linewidth]{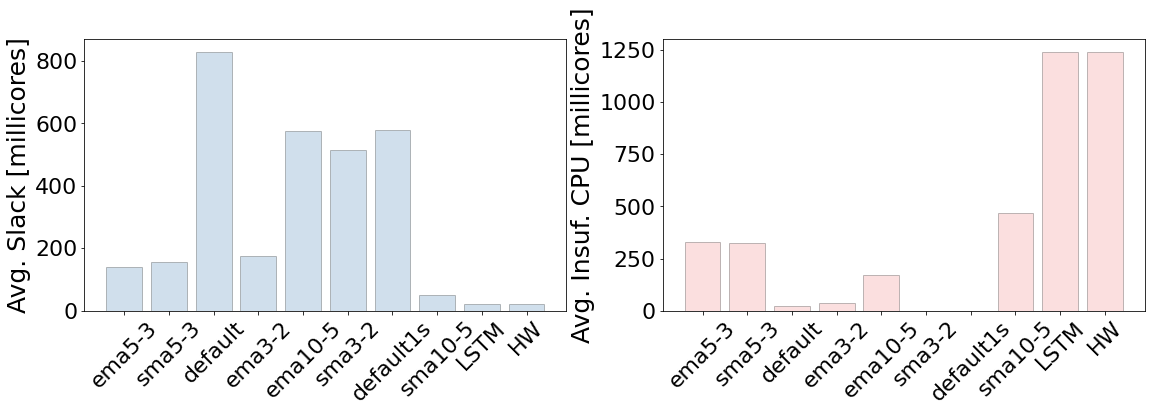}
   \vspace*{-0.4cm}
    \caption{Comparing the average slack and insufficient CPU among all the methods in this paper when running video\_processing\_17m workload in the cold start. Due to the training process of LSTM and HW, they do not perform well in both slack and insufficient CPU as a result of presenting a preset value.} 
    \label{fig:cold_video_bar}
    \vspace*{-0.6cm}
\end{figure}

We experimented with all applications and all autoscalers dynamically allocating CPU for cold application starts. We apply ML-based autoscalers, LSTM and HW using the same data on CPU usage as in the ema5-3 run for a more equitable performance comparison between EMA-based autoscaler and them. We present the behavior under cold starts for the image\_rotate\_shorter and video\_processing\_17m workloads in Figures~\ref{fig:cold_image} and \ref{fig:cold_video}. These figures show that basically the ML-based autoscalers, LSTM and HW, cannot allocate CPU dynamically for cold starts. This is because they do not have any historic information on which the methods could have been trained. Instead, they simply offer a default static amount of CPU. The EMA method we tuned in our previous experiment is able to match the CPU demand accurately.


For all the autoscalers we have implemented for this paper, we show their average slack and insufficient CPU for the cold starts of the two workloads, image\_rotate\_shorter and video\_processing\_17m, in Figures~\ref{fig:cold_image_bar} and \ref{fig:cold_video_bar} correspondingly. ML-based autoscalers present high average insufficient CPU on two workloads (especially in Figure~\ref{fig:cold_video_bar}). It is immediately clear that the ML-based autoscalers cannot follow the CPU demand of the workloads during cold starts due to insufficient historical information. Similarly for the default Kubernetes VPA autoscaler, either significantly under-provisioning or over-provisioning for certain applications, presenting high average slack or average insufficient CPU in Figures~\ref{fig:cold_image_bar} and \ref{fig:cold_video_bar}. Tiny autoscalers achieve extremely low average insufficient CPU but also relatively low average slack at the same time.
\\

\noindent\fbox{%
    \parbox{\linewidth}{%
        \textbf{Conclusion-3:} For cold function starts, ML-based autoscaling algorithms achieve poor performance as they do not have sufficient data to be trained with. The tiny autoscalers can follow the CPU resource demand curves more closely, offering better performance (Figures~\ref{fig:cold_image}, \ref{fig:cold_image_bar}, \ref{fig:cold_video}, \ref{fig:cold_video_bar}).
    }%
}

\subsection{Dynamic CPU Allocation for Warm Starts}

Microsoft shows that in their serverless system~\cite{shahrad2020serverless}, many functions are invoked several times every hour and significant numbers of functions are actually invoked every few minutes. In these conditions, autoscalers that keep track of history have sufficient data to perform well in subsequent functions runs, or warm starts. In this section we investigate how the autoscalers perform when functions are invoked repeatedly, and whether historical information can help in taking better resource allocation decisions.

We run all the workloads under all autoscalers repeatedly for a period of tens of minutes. The first run emulates a cold start, while subsequent runs emulate further warm starts. Using this method we evaluate how the investigated autoscalers perform under warm starts and how accurately they can dynamically allocate CPU to applications.

\begin{figure}[tp]
    \centering
    \includegraphics[width=0.92\linewidth]{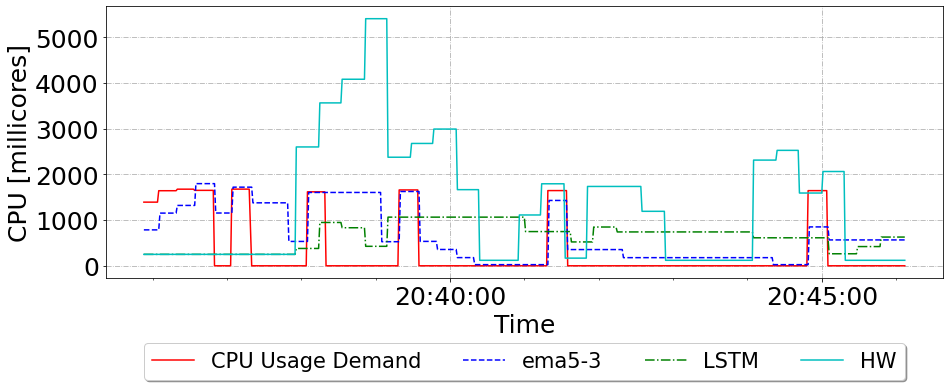}
    \vspace*{-0.3cm}
    \caption{Comparing HW and LSTM with our ema5-3 running on video\_processing\_17m workload. Notice our method follows the CPU usage trend more closely than HW and LSTM, offering better performance.}
    \label{fig:comparison17m_curve}
    \vspace*{-0.2cm}
\end{figure}

Figure~\ref{fig:comparison17m_curve} shows the curves of the performance on the video processing workload for the EMA-based method, HW, and LSTM methods. From Figure~\ref{fig:comparison17m_curve}, the curves indicate that the EMA-based method can follow the trend of CPU usage very well as opposed to methods based on HW and LSTM. Unlike HW and LSTM methods, the EMA-based method does not need data to train the model at the beginning. 

\begin{figure}[tp]
    \centering
    \includegraphics[width=0.92\linewidth]{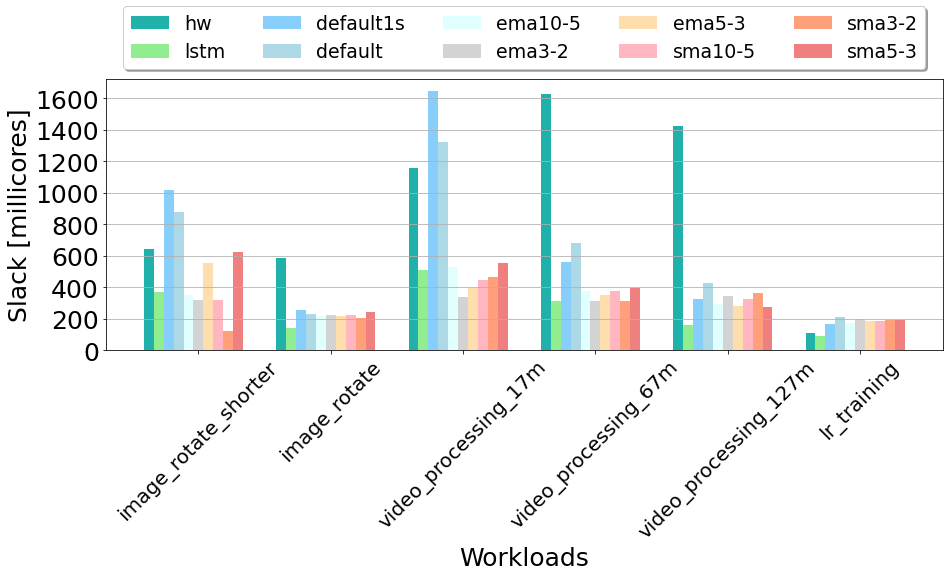}
    \vspace*{-0.45cm}
    \caption{Average CPU slack for all workloads under all autoscalers. The applications were run several times to emulate serverless warm starts.}
    \label{fig:slack_group}
    \vspace*{-0.6cm}
\end{figure}

\begin{figure}[tp]
    \centering
    \includegraphics[width=0.92\linewidth]{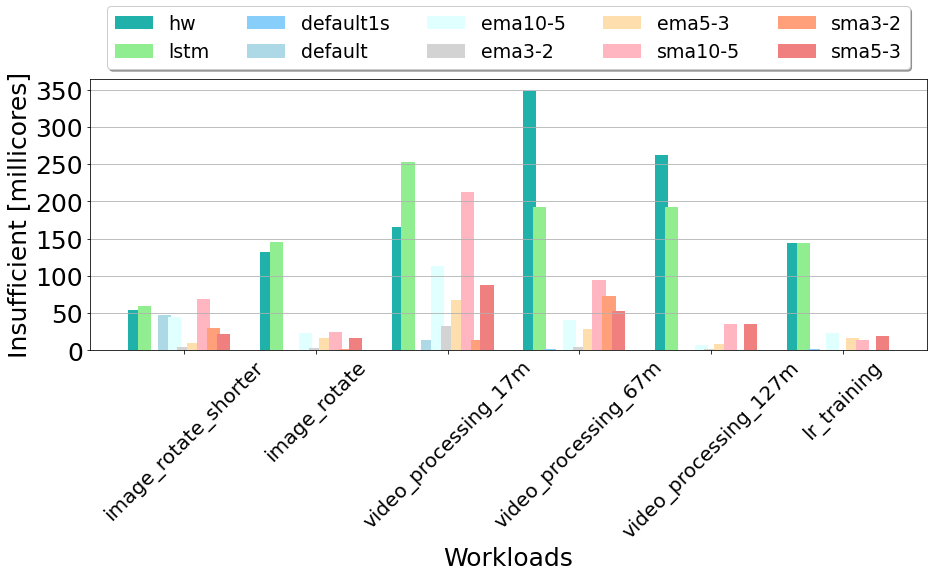}
    \vspace*{-0.45cm}
    \caption{Average insufficient CPU for all workloads under all autoscalers. The applications were run several times to emulate serverless warm starts.}
    \label{fig:insuffi_group}
    \vspace*{-0.6cm}
\end{figure}

We summarize the performance of all the the autoscalers over all applications for warm starts in Figures~\ref{fig:slack_group} and \ref{fig:insuffi_group}. The former presents the average slack (i.e., over-provisioning) and the latter presents the average insufficient CPU (i.e., under-provisioning). For all applications, the best performing autoscalers are the SMA and EMA-based ones that we have tuned. While the default Kubernetes VPA and LSTM and HW autoscalers present extreme behavior---in some situation very high slack and very low insufficient CPU---the EMA and SMA based autoscalers are able to keep both metrics relatively low. This is important because it shows conservative behavior that can help cloud practitioners offer good performance to clients without greatly over-provisioning. Conversely, the amount of application throttles is also kept to a minimum.
\\

\noindent\fbox{%
    \parbox{\linewidth}{%
        \textbf{Conclusion-4:} For warm function starts, tiny autoscalers offer better performance than ML-based autoscaling, following the CPU resource usage more closely and offering less average slack and insufficient CPU usage (Figures~\ref{fig:comparison17m_curve}, \ref{fig:slack_group}, \ref{fig:insuffi_group}).
    }%
}

\subsection{Running Thousands of Tiny Autoscalers}

In practice, to achieve dynamic CPU allocation in serverless clouds, providers have to attach an autoscaler to every running function instance. Recent literature~\cite{agache2020firecracker} shows that providers can run up to thousands of functions per server. In this experiment we show what the added CPU utilization overhead is for running thousands of tiny autoscalers per machine, one for each function.

To only measure the CPU overhead of the autoscaling mechanisms, we create Kubernetes containers that simply sleep after being booted. We then attach to each of these containers a VPA running the EMA5-3 autoscaler. Since the containers are only sleeping, the CPU utilization is only caused by the autoscalers running next to the containers.

\begin{figure}[tp]
    \centering
    \includegraphics[width=0.92\linewidth]{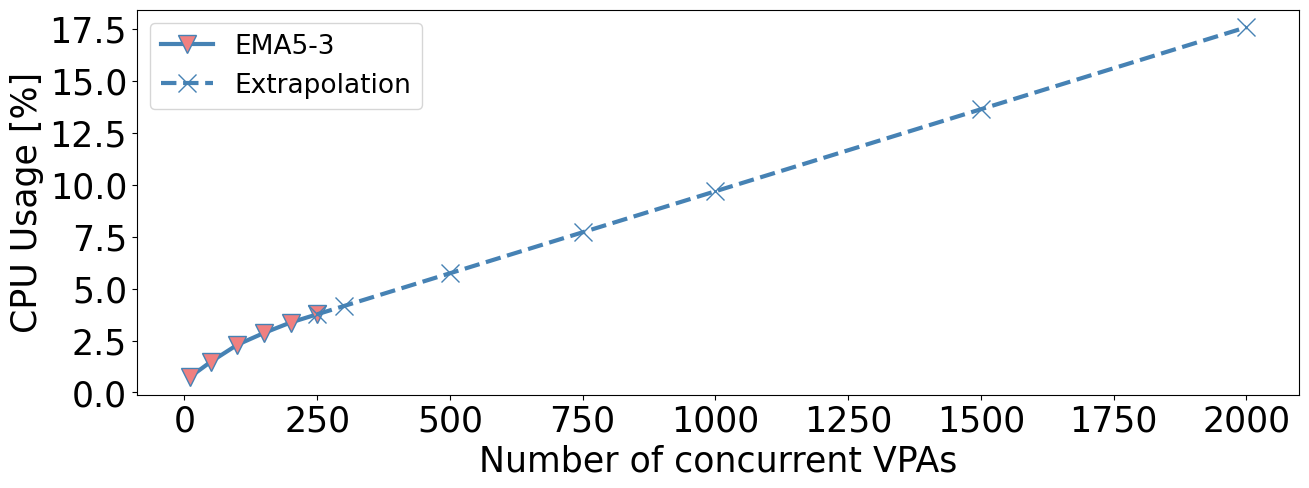}
    \vspace*{-0.4cm}
    \caption{The CPU utilization of running many concurrent EMA5-3 VPAs in Kubernetes. Because Kubernetes cannot run more than 250 VPAs, we extrapolate the CPU usage up to 2000 instances.}
    \label{fig:cpu_usage}
    \vspace*{-0.6cm}
\end{figure}

Officially, Kubernetes developers have put a limit of 500 VPAs running on a single node. Unfortunately, we could  not reach this limit and we could only run reliably up to 250 VPAs. This is not a serverless problem per se, but rather a Kubernetes limitation which does not affect the entire field. In future releases of Kubernetes this issue will be fixed. For this paper, going beyond 250 VPAs, we have used a polynomial curve fitting technique. Our results are plotted in Figure~\ref{fig:cpu_usage}. Following the CPU utilization curve, 250 VPAs do not use more than 5\% CPU. Our extrapolation shows that going up to 2,000 VPAs will not result in more than 17.5\% CPU usage. This leaves sufficient room for the applications to run. If providers consider this limit too high, there are several options to reduce the load: (i) wrapping autoscalers in cgroups with CPU limits; (ii) use a coarser grained monitoring interval, or (iii) group multiple containers (e.g., containers from the same user) under a single autoscaler. \\

\noindent\fbox{%
    \parbox{\linewidth}{%
        \textbf{Conclusion-5:} It is feasible to run thousands of tiny autoscalers, as these fine-grained and lightweight methods do not use much CPU (Figure~\ref{fig:cpu_usage}).
    }%
}

\section{Summary: Is Dynamic CPU Allocation Feasible for Serverless through Tiny Autoscalers?}\label{sec:discussion}

Based on the results we show in Section~\ref{sec:eval}, we return to our main question: is dynamic CPU allocation feasible in serverless environments? This is opposed to common current practice where CPU allocation is fixed and proportional to the memory allocation requested by the client.  

\noindent\textbf{1. Are state-of-the-art autoscalers enough for serverless?}

Currently, the default Kubernetes VPA can not solve the problem of recommendations for sudden and short-lived CPU usage increases very well. The recommendations from the default VPA usually have significant slack, which results in resource under-utilization. The default VPA was designed with different goals, such as longer-running workloads, which it is able to serve well as shown by previous work. Moreover, the ML-based autoscalers, such as LSTM and HW need significant amounts of training data to perform well. Such data might not be available for serverless workloads or might be prohibitively expensive to store at such a large scale.

\noindent\textbf{2. What is a good tiny autoscaler for serverless?} 

For this work we modified the original SMA- and EMA-based autoscalers to slightly over-provision, but within certain bounds, instead of under-estimating. Insufficient resources will lead to throttling of the applications. Conversely, offering many more resources than needed will lead to the under-utilization of the server operated by the cloud provider. 

As our experiments show, the SMA and EMA-based approaches outperform both the ML-based autoscaling, as well as the default Kubernetes VPA. We achieve this result using two techniques. First, we adopt a \emph{bottoming} mechanism for the future CPU demands prediction to prevent it from a rapid decrease. 
Second, we apply the same converging lower bound and upper bound as the default VPA. The lightweight SMA and EMA methods are more appropriate for serverless functions not only in cold invokes but also in warm runs. 



\noindent\textbf{3. Is dynamic CPU allocation feasible for serverless?}

Currently, serverless providers offer clients only statically allocated CPU for running serverless functions. In this work we have investigated whether it is feasible to dynamically allocate CPU to serverless workloads, using tiny autoscalers. Based on our results, we are confident that simple and lightweight techniques, such as SMA and EMA are accurate in predicting and following the CPU utilization of serverless functions. Moreover, practitioners can run thousands of these tiny autoscalers without much CPU overhead. Implementing these in practice might offer serverless platforms the ability to better allocate resources to their clients by reducing overall server under-utilization and by reducing the throttling created by statically allocating CPU.

%% file: relwork.tex
\section{Related Work}
We discuss related work in four categories: prediction of workloads, autoscaling in the cloud, autoscaling containers, and vertical autoscaling virtual machines. 

\textbf{Predicting Workload Trends. } Recent work~\cite{autoscaling-kth} on CPU usage prediction for autoscaling is based on Holt-Winters exponential smoothing (HW) and Long Short-Term Memory (LSTM) methods. These two methods exhibit expensive computational complexity. We have analyzed them in detail in Section~\ref{sec:autoscaling}. The lightweight prediction methods we modify in this article are based on the two-step CPU usage prediction model~\cite{load-prediction}. Similar to these techniques, Casolari and Andreolini~\cite{trend-aware} proposed another trend-aware regression model using linear extrapolation. Regression-based methods~\cite{Regression-Based} also exist in this space, as well as techniques based on autoregression, introduced by Roy et al.~\cite{ARMA}. 

\textbf{Cloud Autoscaling.} A complete taxonomy of the field of cloud autoscaling are developed by Chen at al.~\cite{A-Survey-and-Taxonomy}. Going in depth, several projects specifically investigate the performance study of the state-of-art autoscalers. Versluis et al.~\cite{versluis2018trace}, Ilyushkin et al.~\cite{An-Experimental-Performance} and Jindal et al.~\cite{tool} demonstrate in-depth comparisons for autoscalers on workflows and introduce frameworks and tools to assess the performance of autoscalers. Efficient techniques~\cite{Optimal-Autoscaling} for achieving autoscaling include task allocation strategies, e.g., as the one introduced by Zhong and Buyya~\cite{Cost-Efficient} or by Thurgood and Lennon~\cite{FOSS}, who offered a software solution to autoscale entire Kubernetes clusters. Recently, also serverless clouds have been the subject of autoscaling via reinforcement learning~\cite{schuler2021ai}. However, this only autoscales horizontally the number of container running hosts, not the containers themselves. 


\textbf{Autoscaling Containers.} Rattihalli et al.~\cite{RUBAS} designed RUBAS, an autoscaling mechanism to estimate the CPU and memory resources of containers through the sum of the median of observations and the absolute deviation of observations~\cite{RUBAS-1,RUBAS-2}. Autopilot~\cite{autopilot} introduced by Google is a complete autoscaling system, it not only takes into account vertical autoscaling but also horizontal autoscaling on both CPU and memory usage. The current vertical pod autoscaling recommender we compare against in this article is directly inspired by the moving window recommenders in Autopilot. Nguyen et al.~\cite{hpa} investigated the horizontal pod autoscaling and offered optimization strategies for horizontal pod autoscaling.

\textbf{VM Vertical Autoscaling.} Similar techniques have long been applied to virtual machines in the cloud. These can be vertically either through hotplugging (for CPU or memory) or through CPU throttling (e.g., using rate limiting mechanisms). For example, several articles consider vertical VM autoscaling~\cite{Adaptation,Re-packing-Approach,lu2014application,svard2014hecatonchire}.

As opposed to all these, in this article we assess the feasibility of tiny autoscalers for short-lived serverless functions. Our results show that for these types of workloads, a special kind of autoscaler is needed, namely a lightweight autoscaler that is able to react quickly to changes in demand.

%% file: concl.tex
\section{Conclusions}
In this paper, we have addressed the problem of dynamically allocating CPU for serverless functions through tiny autoscalers. In modern serverless clouds, users request serverless functions of a certain memory size allocation. Subsequently, their CPU is allocated by the cloud provider proportionally to the memory allocation. It is therefore non-trivial for the user to achieve a satisfactory \emph{CPU and memory allocation}. To solve this problem and scale CPU and memory independently, the literature offers various autoscaling algorithms. However, most of them are either heavy-weight or depend on expensive historical information, which may not be available for the short-running and infrequently invoked serverless functions. We therefore investigate five different algorithms for the dynamic rightsizing of serverless functions' CPU during runtime. We implement several of these in Kubernetes and experiment with state-of-the-art serverless workloads. We show that dynamic CPU rightsizing is possible for serverless functions and several algorithms achieve good performance for both cold invokes as well as warm runs. Tiny autoscalers can be found on github: 
\noindent\fbox{%
    \parbox{\linewidth}{
\url{https://github.com/ZhaoNeil/On-Demand-Resizing.git}
}}